\documentclass[prd,preprintnumbers,twocolumn,floatfix,nofootinbib,a4paper,superscriptaddress]{revtex4-2}
\usepackage[normalem]{ulem}
\usepackage{graphicx}
\usepackage{dcolumn}
\usepackage{bm}
\usepackage{macros}
\usepackage{amsmath}
\usepackage{amsfonts}
\usepackage{slashed}
\usepackage{xcolor}
\usepackage[
    colorlinks=true,
    linkcolor=blue,
    citecolor=red,
    urlcolor=red
]{hyperref}

\def\lapp{\mathrel{\rlap{\raise.5ex\hbox{$<$}}
                    {\lower.5ex\hbox{$\sim$}}}}
\def\gapp{\mathrel{\rlap{\raise.5ex\hbox{$>$}}
                    {\lower.5ex\hbox{$\sim$}}}}

\begin{document}


\title{Inelastic Dark Matter from a Higher-Dimensional Operator: The LUX-ZEPLIN High-Recoil Event and Thermal Relic}

\author{Debajyoti Choudhury}
\email{debchou.physics@gmail.com}
\affiliation{Department of Physics and Astrophysics, University of Delhi, Delhi-110007, India.}
\author{Jaydeb Das}
\email{jaydebphys@rnd.iitg.ac.in}
\affiliation{Indian Institute of Technology Guwahati, North Guwahati, Assam-781039,
India.}
\author{Divya Sachdeva}
\email{divyasachdeva@phy.iith.ac.in}
\affiliation{Indian Institute of Technology Hyderabad, Kandi, Sangareddy-502285, Telangana, India}


\begin{abstract}
The recent LUX-ZEPLIN (LZ) data shows a nuclear-recoil event around
$248\,\mathrm{keV}$, motivating scenarios in which dark matter (DM) preferentially scatters at large recoil energies. We
investigate an inelastic scattering interpretation in terms of a simple extension with a fermionic DM
  field interacting with the Standard Model through a single
  dimension-6 operator. A tiny soft symmetry breaking term facilitates
  an endothermic transition while naturally suppressing elastic
  scatterings.  Without any fine tuning, a simultaneous explanation of
  the LZ data and the observed relic density can be effected for a DM
  mass in the $\sim (0.2,1)$~TeV range.  
\end{abstract}
\maketitle

\section{Introduction}
Cosmological observations, most notably measurements of the cosmic
microwave background (CMB), provide compelling evidence that
approximately one quarter of the present energy density of the Universe
is in the form of cold dark matter (DM). Among the earliest and most
extensively studied particle candidates are weakly interacting massive
particles (WIMPs), which can be searched for through nuclear recoils
induced by their scattering in terrestrial detectors.

The LUX-ZEPLIN (LZ) experiment~\cite{LZ:2026axp} has recently reported
an intriguing high-energy nuclear-recoil event with a reconstructed
recoil energy of approximately $E_R\simeq248~{\rm keV}$. For a xenon
target, this corresponds to a momentum transfer
\begin{equation}
q=\sqrt{2m_{\rm Xe}E_R}\simeq250~{\rm MeV},
\end{equation}
which is considerably larger than the momentum transfers typically
probed in conventional WIMP direct-detection searches. An interpretation
in terms of standard elastic scattering of non-relativistic halo DM is
challenging. Producing such a large recoil generally requires particles
from the high-velocity tail of the Galactic DM distribution, while the
nuclear form factor suppresses the scattering rate at large momentum
transfer. Moreover, for the usual contact interactions, the recoil
spectrum decreases towards large $E_R$, making it difficult to
accommodate an isolated event at such a high recoil energy without
simultaneously predicting more events at lower energies. The unusual
recoil energy therefore motivates scenarios in which the scattering
kinematics or the momentum dependence of the DM--nucleus interaction
differs from that of conventional elastic WIMP scattering.

Inelastic dark matter provides a particularly simple possibility. In this scenario, the incoming DM state scatters into a nearby state separated by a small mass splitting $\Delta m$. For the endothermic transition
$\chi_1{\cal N}\rightarrow\chi_2{\cal N}$, with $m_{\chi_2}-m_{\chi_1}
=\Delta m>0$, the minimum incoming velocity required to produce a
nuclear recoil $E_R$ is
\begin{equation}
v_{\rm min}(E_R)
=
\frac{1}{\sqrt{2m_{\cal N}E_R}}
\left(
\frac{m_{\cal N}E_R}{\mu_{\cal N}}
+\Delta m
\right),
\label{eq:vmin_iDM}
\end{equation}
where $\mu_{\cal N}$ is the DM--nucleus reduced mass. Unlike elastic
scattering, $v_{\rm min}(E_R)$ is not monotonic in the recoil energy,
but has a minimum at
\begin{equation}
E_R^\star = m_{\cal N}^{-1} \mu_{\cal N}\Delta m , \quad {\rm with} \quad
v_{\rm min}^{\star}
= \sqrt{2 \mu_{\cal N}^{-1} \Delta m} \ .
\label{eq:ERstar}
\end{equation}

The available halo phase space can therefore be concentrated around a
finite recoil energy. Requiring this 
scale to lie near
the LZ event, $E_R^\star\simeq248~{\rm keV}$, suggests
\begin{equation}
\Delta m\simeq248~{\rm keV}\, \times (m_{\cal N}/\mu_{\cal N})
\end{equation}
corresponding to mass splittings of a few hundred keV over the DM mass
range of interest. A positive $\Delta m$ simultaneously selects the
high-velocity component of the halo and suppresses the lower-recoil
part of the spectrum. Endothermic scattering can therefore shift the
event rate towards relatively large recoil energies, providing a
kinematic mechanism for accommodating the $248~{\rm keV}$ event while
avoiding an accompanying excess at lower recoil energies. Recent
studies have explored fermionic realizations of this inelastic
interpretation and their connections to the dark matter relic
abundance and complementary experimental
constraints~\cite{Fan:2026kxx, Langhoff:2026ujr, *Yin:2026jnn,
  *Wu:2026nhi, *Freese:2026sga, *DiMauro:2026ldr, *Du:2026guj,
  *Visinelli:2026kgt, *Ahmed:2026qjg, *Nomura:2026qyq, *Wang:2026ytg,
  *Bandyopadhyay:2026gjw, *Elahi:2026vlm, *Yamashita:2026ump,
  *Jeesun:2026vzo, *Unwin:2026rdp, *Smirnov:2026aqk, *deLima:2026shq,
  *Lee:2026wof, *Das:2026uyy, *Liang:2026coz, *Kannike:2026qyl,
  *Du:2026lpa, *Alhazmi:2026efz, *Okada:2026eol, *Lee:2026xxh,
  *Asadi:2026iot, *Zhu:2026dag, *Yuan:2026djt, *Kumar:2026lgi,
  *He:2026hqz, *Heikinheimo:2026kwp, *Su:2026rwz, *McCabe:2026crm,
  *Gu:2026vto, *Khan:2026nwp, *Chattaraj:2026fxn, *Okada:2026upm,
  *DiMauro:2026ymt, *He:2026idw, *Uttayarat:2026isp,
  *Palmisano:2026kuj, *Baer:2026yrt, *Ahmed:2026kan, *Lian:2026hpm,
  *Das:2026buc, *Barman:2026omh, *Borah:2026ris, *Bose:2026szs,
  *Khan:2026osp, *Gemmell:2026yaw, *Fan:2026hzw, *Nagata:2026pbj}.

In this work, we explore this possibility in a simple and economic
effective field theory (EFT) framework in which a Dirac fermion $\chi$
interacts with the Standard Model (SM) through a single dimension-six operator.
The scenario naturally reproduces the kinematic features of the 
LZ event while maintaining consistency with other direct detection
experiments. Moreover, the very same operator realizes the correct relic
abundance without any fine tuning whatsoever.

\section{EFT FRAMEWORK}

We extend the Standard Model (SM) by a gauge-singlet Dirac fermion
$\chi$. Rather than introduce model-specific messengers, we assume
that its interaction with the SM fields is governed by a single
dimension-6 operator also involving the Higgs field $H$ and its
covariant derivative.  Thus,
\begin{equation}
\mathcal L_\chi
= \bar\chi(i\slashed{\partial}-m_\chi)\chi
+ (C_{V}/\Lambda^2) (\bar\chi\gamma^\mu \chi) (H^\dagger i\overleftrightarrow D_\mu H),
\label{eq:lagr}
\end{equation}
where $\Lambda$ is the cutoff scale of the EFT and $C_V$ is a real
Wilson coefficient. While other operators (both dimension-5 and
dimension-6) are possible, our choice is dictated by the twin
 imperatives of the inelastic event at LZ and
the lack of elastic events at direct detection experiments. Our
approach is phenomenological and we remain agnostic as to the identity
of the UV-completion that preferentially chooses $C_V$ over others.

The Lagrangian is invariant under a global $U(1)$ under which
only $\chi$ has a non-zero charge. Augmenting $\mathcal L_\chi$
by a soft term
\begin{equation}
\mathcal L_\chi \to \mathcal L_\chi +\frac{m'}{2}
\left(\overline{\chi^c}\chi+\mathrm{h.c.}\right)
\end{equation}
breaks the $U(1)$ down to an unbroken $Z_2$, and splits the Dirac fermion into two Majorana states,
{\em viz.}
\begin{equation}
\chi_1=\frac{1}{\sqrt{2}}\left(\chi+\chi^c\right),
\qquad
\chi_2=-\frac{i}{\sqrt{2}}\left(\chi-\chi^c\right),
\end{equation}
with masses
\begin{equation}
m_{\chi_1}=m_\chi-m',\qquad
m_{\chi_2}=m_\chi+m'.
\end{equation}
Without any loss of generality, we define $\chi_1$ to be the lighter
state (and the putative DM) with the mass splitting between the states
given by $\Delta m\equiv m_{\chi_2}-m_{\chi_1}=2m'$.

The $\chi$-current, then, can be rewritten as
\begin{equation}
  \bar \chi \gamma_\mu \chi
  =   \frac{i}{2}\left(\bar \chi_1 \gamma_\mu \chi_2 -   \bar \chi_2 \gamma_\mu \chi_1\right) =i \bar \chi_1 \gamma_\mu \chi_2
\label{eq:chi_current}
\end{equation}
and it is this off-diagonal nature that would facilitate the inelastic
 scattering process
\begin{equation}
\chi_1 + {\cal N}\rightarrow \chi_2 +{\cal N},
\label{eq:inel_scatt}
\end{equation}
leading to the nuclear-recoil signal 
at LZ, while suppressing the elastic scattering. On a similar vein,
the relic abundance (within the freeze-out paradigm) would be ensured by
co-annihilation rather than annihilation.

\section{Relic density and Other Constraints}
Post electroweak symmetry breaking, the Higgs doublet can be written
as $H = \left(0, \; (v + h)/\sqrt{2}\right)^T$ and, thus, 
  \begin{equation}
\mathcal L_\chi
\supset
\frac{-g}{2c_W}
\frac{C_V}{\Lambda^2}
\left(v^2+2vh+h^2\right)
Z_\mu\,\bar\chi\gamma^\mu\chi.
\label{eq:eft_after_ewsb}
\end{equation}
with the $\chi$-current given by eqn.(\ref{eq:chi_current}).  Thus, in
addition to the $Z\chi\chi$ interaction, the operator also generates
$Zh\chi\chi$ and $Zhh\chi\chi$ interactions.

The smallness of $C_V v^2 / \Lambda^2$ ensures that the
  constraints from the invisible decay width of the $Z$ would be of
  little concern even if $Z \to \chi_1 \chi_2$ were kinematically
  accessible. Of cosmological concern could be the decay modes 
\begin{equation}
  \chi_2\rightarrow\chi_1+\nu+\bar\nu \quad {\rm and} \quad
  \chi_2 \to \chi_1 + n \gamma,
\end{equation}
where the first (and dominant) mode proceeds through an off-shell $Z$ boson
while the second is loop driven. The corresponding width is
\begin{align}
    \Gamma\,&\simeq\,\frac{1}{40\pi^3}\left(\frac{C_{V}}{\Lambda^2}\right)^2{(\Delta m)^5}.
\end{align}
For the parameter space of interest, the lifetime is much shorter than
the age of the Universe, and the heavier state therefore decays away
well before the present epoch. Since the decay converts $\chi_2$ into
the stable state $\chi_1$, the total dark matter number is
approximately preserved, while the small energy associated with the
mass splitting is transferred to the neutrino (photon) sector. We
therefore treat the present-day dark matter population as consisting
entirely of the lighter state $\chi_1$. However, with much of the
decay occurring post-Big Bang Nucleosynthesis (although mostly before
the CMB epoch), naively, constraints from these sectors could be
important.  Note, though, that the small mass splitting implies that
not only there is no hadron production, but the energy released to the
photons is too small to initiate photodisintegration of light
elements~\cite{Lee:2026jxl}. Similarly, the energy injected into
neutrinos is too small to be relevant cosmologically.

Turning to relic abundance, the DM freezes out after the EWSB epoch.
With $\Delta m \ll T_f \sim m_\chi/20$, the post-freeeze out
abundances of $\chi_{1,2}$ are nearly equal. The depletion thereafter
is primarily governed by the coannihilation rates for
\begin{equation}
  \chi_1\chi_2\rightarrow Zh, \, W^+ W^-, \, f\bar f,
  \label{eq:coannihil_modes}
\end{equation}
with the first one (driven by the contact term while the others are
$s$-channel $Z$-exchange) dominating if kinematically
accessible. {The relevant thermal average, then is
  $\langle v \sigma_{12} \rangle$---with $\sigma_{12}$ being the total
  cross section for the processes in eqn.(\ref{eq:coannihil_modes})---and is computed using \texttt{micrOMEGAs} \cite{Belanger:2004yn,Alguero:2023zol}. The
  smallness of $\Delta m /T_f$ also ensures that the
  relic abundance is insensitive to it.}

\begin{figure}
    \centering
    \includegraphics[width=0.9\linewidth]{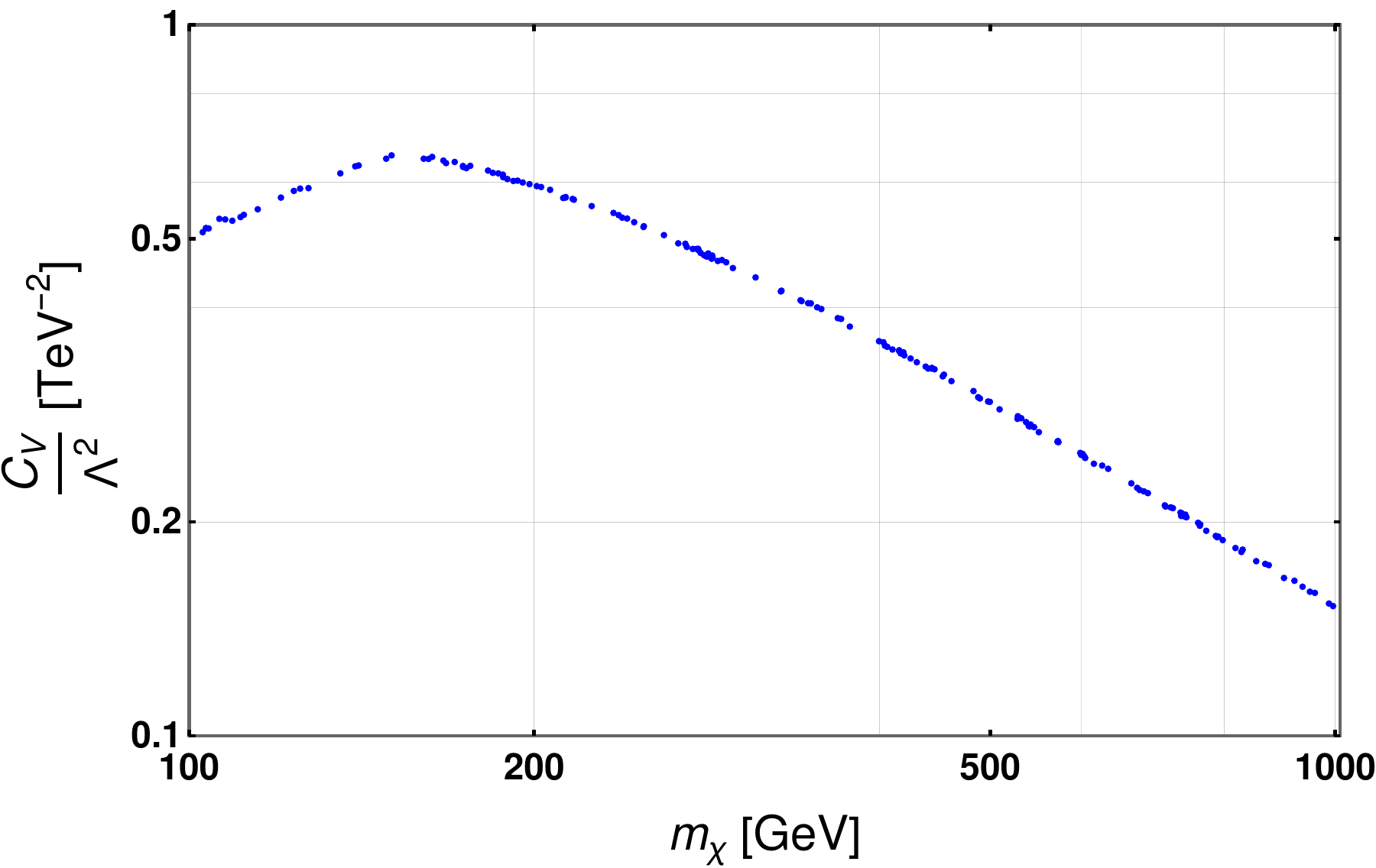}
    \caption{$C_{V}/\Lambda^2$ as a function $m_{\chi}$, subject to $0.119\leq\Omega_{\chi}h^2\leq0.121$~\cite{Planck:2018vyg}.}
    \label{fig:relic}
\end{figure}

As Fig.\,\ref{fig:relic} shows, only very moderate values of
$C_V/\Lambda^2$ are required to obtain the correct relic density.
The relic-density condition requires an effective annihilation rate that varies only weakly with the DM mass. The required coupling nevertheless depends on the mass dependence of the annihilation amplitude square and the available final-state phase space. At lower DM masses, the finite \(W\), \(Z\), and Higgs masses suppress the available phase space for the \(W^+W^-\) and \(Zh\) final states, requiring a larger coupling to obtain the observed relic abundance. This is reflected by the behaviour of the curve
in the low $m_\chi$ range where, in addition, the cross-sections
suffer from kinematic suppressions owing the SM particle masses.  For
large $m_\chi$, the cross-sections scale as $(C_V m_\chi /
\Lambda^2)^2$ and this explains the slow fall of $(C_V /\Lambda^2)$
with $m_\chi$.

\section{The LZ event}
The differential nuclear-recoil rate for the inelastic process
$\chi_1{\cal N}\rightarrow\chi_2{\cal N}$ is given by
\begin{equation}
\frac{dR}{dE_R}
=
\frac{\rho_\chi}{m_\chi}\sum_A \frac{\xi_A}{m_A}
\int_{v>v_{\rm min,A}(E_R)}
d^3v\,
f_{\rm lab}(\mathbf v)\,
v\,
\frac{d\sigma_{\cal A}}{dE_R},
\label{eq:iDM_rate}
\end{equation}
where $\xi_A$ is the mass fraction of isotope $A$ in the detector, $m_A$ is the mass of the corresponding target nucleus, $f_{\rm lab}(\mathbf v)$ is the DM velocity distribution in the
laboratory frame, and we adopt
the Standard Halo Model (SHM)~\cite{McCabe:2010zh}, with
$\rho_\chi=0.4~{\rm GeV\,cm^{-3}}$, $v_0=220~{\rm km\,s^{-1}}$,
$v_{\rm esc}=544~{\rm km\,s^{-1}}$, and $v_E=232~{\rm
  km\,s^{-1}}$. The Galactic-frame Maxwellian distribution is
truncated at $v_{\rm esc}$ and boosted to the laboratory frame, giving
a maximal laboratory-frame speed $v_{\rm max}=v_{\rm
  esc}+v_E\simeq776~{\rm km\,s^{-1}}$.  For the coherent nuclear
response, we  evaluate the
recoil spectrum (using the Helm form factor) separately for each
naturally occurring Xenon isotope, combining
the individual contributions according to their natural isotopic
abundances.  Finally, we fold the differential rate with the
nuclear-recoil acceptance $\epsilon(E_R)$~\cite{LZ:2026axp}
and an energy-response function, {\em viz.}
\begin{equation}
\frac{dN}{dE_{\rm obs}}
=
{\cal E}
\int dE_R\,
G(E_{\rm obs};E_R,\sigma_E)\,
\epsilon(E_R)\,
\frac{dR}{dE_R},
\label{eq:LZ_detector}
\end{equation}
where ${\cal E}=2.84~{\rm tonne}$-${\rm yr}$ is the exposure.  We model the
reconstructed recoil energy using a normalized Gaussian,
\[
G(E_{\rm obs};E_R,\sigma_E)
=
\frac{1}{\sqrt{2\pi}\sigma_E(E_R)}
\exp\left[
-\frac{(E_{\rm obs}-E_R)^2}
{2\sigma_E^2(E_R)}
\right] \ ,
\]
with the
energy-dependent width being~\cite{Fan:2026hzw}
\[
\sigma_E(E_R)
=
\sqrt{
(23~{\rm keV})^2
\left(\frac{E_R}{248~{\rm keV}}\right)
+
(0.093\,E_R)^2
} \ ,
\]
thereby reproducing the reported
energy uncertainty,
\begin{equation}
  E_R=248\pm23_{\rm stat}\pm23_{\rm sys}~{\rm keV}.
  \label{eq:LZ_bin}
\end{equation}

To understand the LZ signal profile, it is
conventional to consider a reference DM-neutron $(n)$ cross
section~\cite{Essig:2007az,Rodd:2026tyn}, which, for our case is
defined as
\begin{equation}
\sigma_n^{\rm ref}
= (\mu_{\chi n}^2/4\pi) \times (C_V/\Lambda^2)^2.
\label{eq:sigmaN}
\end{equation}

While $\sigma_n^{\rm ref}$ is essentially
independent of $\Delta m$ for the splittings considered here, the
observable event rate depends strongly on it. 
{Scattering off a nucleus is coherent, and for an isotope $A$ (with $N_A$ neutrons and $Z_A$ protons), the effective cross section is given by}
\[
\sigma_{\rm eff}
\approx
\sigma_n^{\rm ref}\,
\frac{\mu_{\chi A}^2}{\mu_{\chi n}^2}
\left[N_A-(1-4\sin^2\theta_W)Z_A\right]^2
     {\cal K}(m_\chi,\Delta m),
\label{eq:Nsig_schematic}
\]
where $\mu_{\chi A}$ is the DM--nucleus reduced mass and ${\cal K}$
contains the halo integral, nuclear form factor, detector acceptance,
and energy resolution. The dominant dependence on the mass splitting
can be encapsulated in a multiplicative factor $\eta[v_{\rm
    min}(E_R)]$ where
\begin{equation}
\eta(v_{\rm min})
=
\int_{|\mathbf v|>v_{\rm min}}
d^3v\,
\frac{f_{\rm lab}(\mathbf v)}{v},
\label{eq:eta_vmin}
\end{equation}

\begin{figure}[!ht]
    \centering
    \includegraphics[width=\columnwidth]{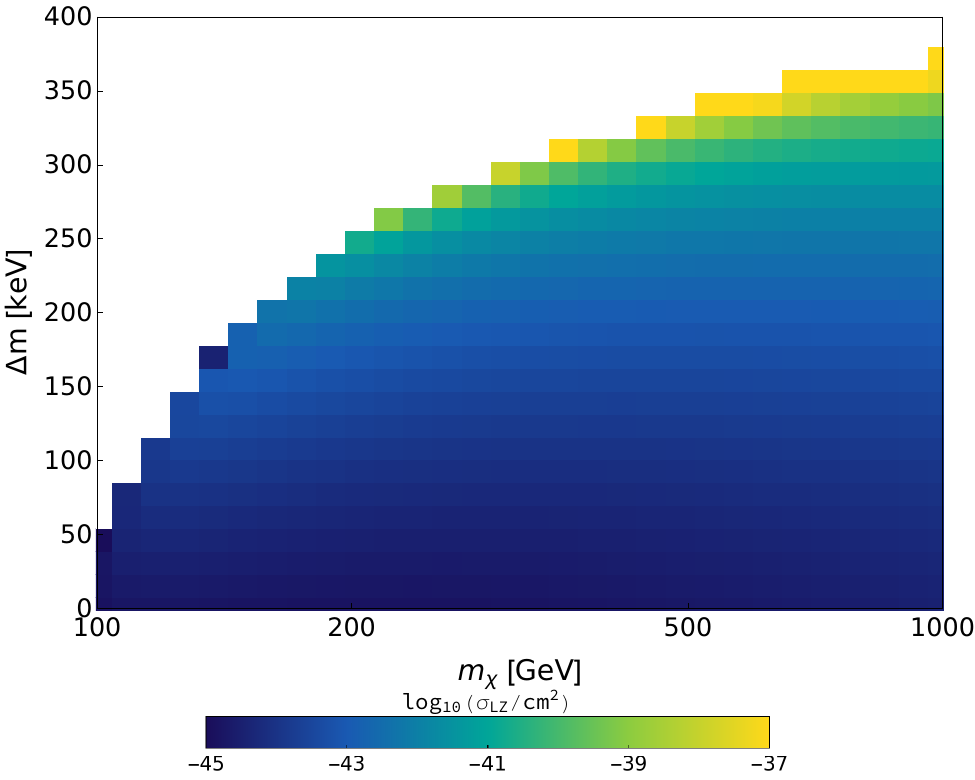}
    \caption{Best-fit reference DM-neutron cross section
      $\sigma_n^{\rm LZ}$ in the $(m_\chi,\Delta m)$ plane, inferred
      from the LZ high-energy event and the absence of events in the
      control interval. The color scale shows $\log_{10}(\sigma_n^{\rm
        LZ}/{\rm cm}^2)$. The result assumes an endothermic contact
      interaction with the weak-charge nuclear response, the Standard
      Halo Model, and the detector response specified in the text, but
      does not depend on a particular ultraviolet completion.}
    \label{fig:LZ_sigma}
\end{figure}

Using the resulting efficiency-weighted and smeared recoil spectrum,
we determine, for each $(m_\chi,\Delta m)$ combination, the reference
cross section (see Fig.\,\ref{fig:LZ_sigma}) required to reproduce the
observed {signal profile. This is
  effected} using a likelihood that includes the observed high-energy
event and the zero-event control interval ($[10\,{\rm keV},225\,{\rm
    keV}]$).  At smaller $\Delta m$, the velocity threshold is lower
  and a larger fraction of the halo population contributes, so a
  smaller cross section is sufficient to produce the observed event.
  Increasing $\Delta m$ raises the velocity threshold and reduces the
  available halo phase space thereby increasing the required
  $\sigma_N^{\rm LZ}$.  For a given $(m_\chi, \Delta m)$ combination
  then, the required $C_V/\Lambda^2$ can be inferred from
  eqn.(\ref{eq:sigmaN}) in a straightforward manner, and the results
  lie in the acceptable domain.

\section{Discussion and Epilogue}

Having established that both the observed relic density as well as the
LZ signal profile (with the attendant negative results at other direct
detection experiments) can be explained in this simple model, it is
important to investigate whether the same set of parameters can yield
both results simultaneously. In other words, can we find
\[
C_V({\rm relic}) = C_V({\rm LZ})
\]
where the two evaluations are made independently from the two
observations?  The mass splitting is key to answering this, through its
role in determining $E_R^\star$ and $v_{\rm min}$ (which, of course,
has to be less than $v_{\rm max}$) and thereby determining, in turn,
$C_V({\rm LZ})$.

\begin{figure}[!ht]
    \centering
    \includegraphics[width=\columnwidth]{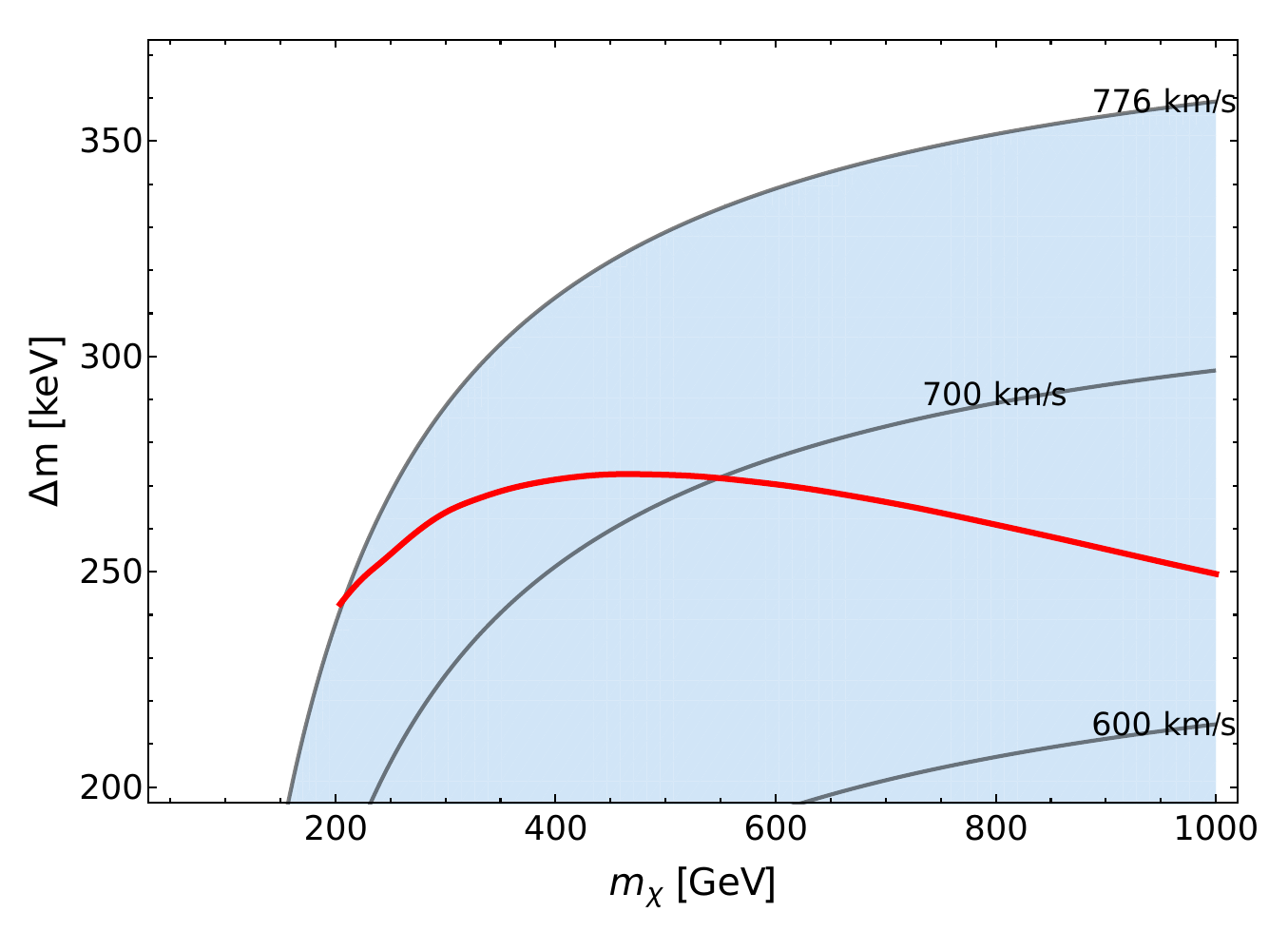}
    \caption{
    Minimum DM velocity required to produce a nuclear recoil of
    $E_R=248~{\rm keV}$ in the $(m_\chi,\Delta m)$ plane.
    The black contours show constant $v_{\min}$, while the red curve
    denotes the parameter values for which both the observed relic abundance and
    the LZ event {can be explained by the same $C_V$.}
    The shaded region is kinematically accessible within the adopted
    Standard Halo Model, with
    $v_{\rm max}=v_{\rm esc}+v_E\simeq776~{\rm km\,s^{-1}}$. 
    The matching solutions preferentially probe the high-velocity tail
    of the Galactic DM distribution while remaining below the
    kinematic boundary.
    }
    \label{fig:vmin_relic_LZ}
\end{figure}
In Fig.\,\ref{fig:vmin_relic_LZ}, we present the relation between
$m_\chi$ and $\Delta m$ that offers a simultaneous solution. Note that
only a relatively small range is acceptable, {\em viz.}  $240 \, {\rm
  keV} \lapp \Delta m \lapp 270 \, {\rm keV}$ is allowed, a result
that is not surprising given that we required the excess to be confined to
a single bin centred around $E_R = 248~{\rm keV}$ (see eqn.\ref{eq:LZ_bin}).

For {a given} DM mass, the {the relic abundance determines} 
$C_V/\Lambda^2$  {which, in turn, determines the total cross section at LZ.}
 The mass splitting is then {chosen so as} to {satisfactorily}
reproduce the LZ recoil signal {profile}.
Where coannihilation is less efficient, including regions affected by
final-state phase-space suppression, the relic abundance requires a
larger coupling and therefore implies a larger scattering cross section.
For a small mass splitting, this normalization can overproduce the recoil
signal; a larger positive $\Delta m$ suppresses the rate by restricting
scattering to the high-velocity tail of the halo distribution. At larger
masses, the growth of the annihilation rate at fixed $C_V/\Lambda^2$
permits a smaller relic-compatible coupling, reducing the scattering
normalization and the kinematic suppression needed to reproduce the
event. The resulting nonmonotonic relation between $m_\chi$ and
$\Delta m$, including the maximum of the red curve, reflects the combined
mass dependence of the relic-compatible cross section and the
detector-weighted inelastic recoil response.

We have, thus, successfully demonstrated that a very simple extension
of the SM, in the form of a single fermionic field and just one
dimension-6 operator connecting it to the SM suffices to explain the
puzzling LZ data as also reproduce the observed relic density. The
model is a very economic one and provides a robust template for future
model building should more details emerge with added exposure time. We
end by reminding of possible caveats. The model is purely
phenomenological in its choice of the Wilson coefficients, the guiding
principle being the null result in elastic scattering data at direct
detection experiments. More such phenomenologically acceptable operators
exist though and switching these on would only serve to change the
quantitative results presented in Fig.\ref{fig:relic} and, thus,
Fig.\ref{fig:vmin_relic_LZ}. We desist from attempting either this or
a UV completion, until more data is accumulated and more features
discernible.

\section*{Acknowledgment}
DC acknowledges the IoE, University of Delhi grant
IoE/2025-26/12/FRP. JD acknowledges the ANRF, Government of India, for
the National Postdoctoral Fellowship grant PDF/2023/0015. DS
acknowledges the ANRF for support through the project
ANRF/ECRG/2025/002846.

\bibliographystyle{apsrev4-2}
\bibliography{Refs}
\end{document}